\documentclass{article}

 \usepackage{setspace}
\usepackage{amssymb,latexsym,amsmath}

\usepackage{graphicx}
\usepackage{rotate}

\begin{document}

\title{Estimating migration proportions from discretely observed continuous diffusion processes}

\author{V.\ Calian\footnote{Corresponding author: V. Calian, Science Institute, University of Iceland, Dunhaga 3, 107 Reykjavik, Iceland, calian@raunvis.hi.is, tel: +354-5254777, fax: +354-552-8911 } and G. Stefansson\\
Science Institute, University of Iceland\\
Dunhaga 3, 107 Reykjavik, Iceland \\
\\
L. P. Folkow, A.S. Blix \\
Department of Arctic Biology, University of Tromso \\ 
Breivika, N-9037, Tromso, Norway\\
}
\date{}

\maketitle

\begin{abstract}
We model two time and space scales discrete observations by using a unique continuous diffusion process with time dependent coefficient. We define new parameters for the large scale model as functions of the small scale distribution cumulants. We use the non - uniform distribution of the observation time intervals to obtain consistent and unbiased estimators for these parameters. Closed form expressions for migration proportions between spatial domains are derived as functions of these parameters. The models are applied to estimate migration patterns from satellite tag data.

\end{abstract}


\newpage

\section{Introduction}

Statistical analysis of ecological complex systems \cite{1}, \cite{2}, financial data \cite{3} or genetics \cite{4} increasingly relies on stochastic models for data underlying processes. In addition, most cases require integration of several types of deterministic and stochastic models \cite{5}.  Presence of errors, with a priori unknown distribution makes estimation even more difficult.

In this paper, we propose a method of statistical inference at long time and large space scales when the available data consists of  discrete observations, measured (with generally distributed errors) at non - identical time intervals and much smaller scale.  For this purpose, we define meaningful process parameters and find corresponding unbiased and consistent estimators which can be used for inference.

We are motivated by a specific type of data, consisting of multiple time series of spatial locations,with finite lengths, measured at unequal, finite time intervals. This is a typical structure for observations of complex ecological systems with migration processes, such as observations from automatic positioning instruments recording location using GPS signals in data storage tags (DSTs).
Aggregated counts on spatially extended domains, at given time intervals may be also available and need to be simultaneously used. In either cases, one is interested in predictions of migration proportions between spatial domains, over large time intervals, for ensembles of possibly non-identical individuals of the given system.

Models of population dynamics quickly become analytically infeasible and
this is why numerical approaches abound, some even with little theoretical
justification.  Detailed multispecies models of population dynamics commonly
need to include spatial structure to describe temporally variable species
overlap \cite{6} and these can quickly become computationally
infeasible.  For example, models with unknown temporally varying migration
rates between several areas give obvious estimation problems,
particularly in a multispecies context.  It is therefore important to
formulate migration in such a manner as to reduce the number of parameters,
yet allow both flexibility and permit the incorporation of the migration
process into typical box models.  In particular, in a complex framework such
as a multispecies, multiarea Gadget model
\cite{7}, \cite{8}, \cite{9}, it is not feasible to incorporate
a computational layer which requires numerical solutions to partial
differential equations to describe migration.  Rather, solutions in closed
form are required to describe the migration processes.

We assume that the {\it observation scale} underlying process is a fairly general diffusion \cite{10}, \cite{11} . The continuous model which is discretely observed may be regarded as the limit of a biased random walk (unobserved, at much smaller scale) with identical / {\it non-identical} steps, i.e. with constant or time - dependent drift and diffusion coefficients, respectively. 
If several spatial paths are observed, we assume same number of  independent diffusions  as underlying processes. 

Diffusion processes may be described in several ways. One can use the stochastic equation representation of the type $d{\bf r}_t = \beta dt + {\bf D} dB_t$, for general drift ($\beta$) and diffusion ($D$) which may depend on time and on the process ${\bf r}_t$, and with $B_t$ a Brownian motion. We will use the complementary representation, the partial differential equation (Kolmogorov - forward or Focker - Planck equation)  which describes the evolution of the probability density function $P({\bf r}, t)$ in time and space:

\begin{equation} \label{FP}
\frac{\partial P({\bf r}, t)}{\partial t} = - \nabla {\bf \beta} P({\bf r}, t) + \frac{1}{2} \nabla^2 {\bf D} P({\bf r}, t)
\end{equation}
Here, ${\bf D} $ is the  diffusion matrix and ${\bf \beta}$ - the drift vector, for a general d - dimensional case (${\bf r} \in {\bf R}^d$). Note that higher order derivative terms could be included in equation (1) when considering more general models. 

The {\it inference scale}, much larger, will be characterized by the same underlying processes but different initial and boundary conditions for the equation (1), imposed by the ecological constraints, in our case. We give (section 2) closed form solutions for the migration proportions, which depend on newly defined large scale drift and diffusion parameters.

Diffusion models have been frequently employed  in modeling migration (\cite{1}  , \cite{2}). Most of them rely on numerical solutions which would at least slow-down considerably any complex system analysis which involves several time scales and several deterministic and stochastic processes. Severe limitations related to such solutions can be avoided by using analytical approximations as provided in \cite{12} for the case of one dimensional diffusion processes and Gaussian noise.

By contrast, we use the non - uniform time - interval distribution as an advantage in calculating the cumulants of the long time large scale distribution of observations as a function of the smaller observation scale. This allows us to introduce what we call {\it effective} and {\it collective} models, parameters and their estimators  (section 3). We illustrate our method with a real data example of migration, in section 4.

\section{Process modeling and main assumptions}

In this section we briefly review two typical solutions of the Focker - Planck equation (1), which will play  key roles in the construction of statistical models in section 3, since they provide the distributions of the true (under the model) values of positions for a given time interval distribution.

Although we allow for time dependence of the diffusion coefficient, we still make a series of simplifying {\bf assumptions}:

\noindent (a) the lengths of time intervals between observations, the measurement errors and the true positions are independently distributed; 

\noindent (b) errors are independently and identically distributed (according to some general, non - Gaussian law). Errors and process are also independent.
 
\noindent (c)  the space - domain $\Omega$ is 2 - dimensional ($d=2$).  
We will actually work only on rectangular domains, in order to give analytical solutions as far as possible. This can be generalized to more general geometries, but keeping closed forms for the results  would require other  assumptions.

\noindent (d) the diffusion is homogeneous in space, the matrix ${\bf D}$ is diagonal with identical elements $D(t)$ which may depend on time.

\noindent (e) the link between the observed (${\bf r}_i^{obs}$) and true (${\bf r}_i$) values is given by a simple additive statistical model: $ {\bf r}_i^{obs} = {\bf r}_i + {\bf} \epsilon_i $, where ${\bf} \epsilon_i$ are measurement errors.


\noindent (f) we define the distribution of errors $\bf \epsilon$ in terms of cumulants $k_a^{\epsilon}$, with $k_1^{\epsilon}= {\bf 0}$, a diagonal $k_2^{\epsilon}$ matrix, and possibly higher order cumulants $k_a^{\epsilon}(\epsilon_{i_1}, ..., \epsilon_{i_a})$. The variance-covariance matrix for the error distribution is assumed to be diagonal and we choose for simplicity ($(k_2^{\epsilon})_{jj}= \sigma_0^2$).

\subsection{Discrete observations scale}


We are motivated by position and time data recorded by satellite tags for migration studies. They provide a large number of observations $ {\bf r}^{obs}_0,..., {\bf r}^{obs}_n$, at finite time - intervals $t_0,..., t_n$, for many finite paths $\gamma \in \Gamma$,  where the set $\Gamma$ is included the spatial domain $ \Omega $. 

\noindent The boundary conditions for equation (1) will be:

\begin{equation} \label{smallBC}
P({\bf r} \rightarrow \pm \infty, t ) \rightarrow 0
\end{equation}
since we assume that the boundary of the spatial domain $\Omega$ is very "far"  from any observed path.

If $D(t)$ is constant in time, the solution of Focker-Planck equation becomes: 
 $P({\bf r}, t)= \int G(\delta {\bf r} \mid \delta t) P_0({\bf r}_0, t_0) d{\bf r}_0 $. Here $P_0({\bf r}_0, t_0)$ are initial conditions and the transition density is a Gaussian in true (under the model) position differences $\delta {\bf r} = {\bf r - r_0} $ corresponding to any given time intervals $\delta t = t-t_0$ :

\begin{equation} \label{smallSol}
G(\delta {\bf r} \mid \delta t)= \frac{1}{ (4 \pi D \delta t )^{d/2} }  \exp \left( - \frac{(\delta {\bf r} - {\bf \beta} \delta t)^2}{4D \delta t}  \right)
\end{equation}


\vspace{0.3cm}
\noindent {\bf Remark 2.1}

\noindent The Green function solution of the Focker - Planck equation with time - dependent diffusion coefficient $D(t)$ is still a Gaussian:
$ \frac{1}{ (4\pi \int D(t) dt)^{d/2} }  \exp \left( - \frac{(\delta {\bf r}_i - \vec \beta \delta t)^2}{4\int D(t)dt}  \right) $. 
This will allow us to solve both  statistical inference problems (for constant and time dependent diffusion) in a very similar manner.

\subsection{Large scale counts and proportions}

The second type of observations we need to model are the counts on extended spatial domains (we will denote coordinates ${\bf R} \in \Omega$ to distinguish from the finer spatial scale), at given (long)  time intervals ($\Delta T$). 
An example is provided again by migration studies (mark - recapture data), where classical tags are used and only aggregated counts can be recorded, at longer time intervals. 

Our main goal is to estimate migration proportions, i.e. the fraction of paths which start in a given spatial domain and end in an other domain, after a given time $\Delta T$. We derive here the theoretical expressions of these proportions, as functions of process parameters and we will show in the next section how these parameters can be estimated.

The same stochastic process is assumed to generate the true values. The same differential equation for the probability distribution function has to be solved, but for different boundary conditions:

\begin{equation} \label{BClarge}
\left( \frac {\partial P({\bf R}, T )}{\partial {\bf R}} \right)_{{\bf R} \in \partial \Omega} =0
\end{equation}
and with
$ P({\bf R}, T_0)= $ constant on a given $\Omega_0$
as initial conditions. Here, the distances $R = ||{\bf R}|| $ are much larger than the typical distances $r=||{\bf r}|| $ in previous section. These are particular conditions we chose  in order to model the fact that the migrating individuals are not leaving a given habitat ($\Omega$), the distances between observations are comparable with the characteristic lengths (denoted $L_x, L_y$) of the domain $\Omega$  and that the time intervals between two counting experiments are large enough for their distribution to become uniform on a given area.
Note that in fact, these initial conditions are also the long time ($\Delta T >> \delta t$)  limit of the solutions of the previous problem (subsection 2.1).

The Green functions solution can be explicitly calculated, for arbitrary $\Delta {\bf R}= \bf R_f - \bf R_i$ (with coordinates $\Delta X, \Delta Y$), under the assumption (b):

\begin{equation} \label{largeSol}
G(\Delta {\bf R} | \Delta T ) = G(\Delta X | \Delta T ) G(\Delta Y | \Delta T ) 
\end{equation}
Here:

\begin{eqnarray} \label{LargeSol}
& G(\Delta X | \Delta T )= \sum_n (I_{1n} + I_{2n}) \\ \nonumber
& G(\Delta Y | \Delta T )= \sum_n (I'_{1n} + I'_{2n})
\end{eqnarray}
and:

\begin{equation}
I_{1n} = \frac{1}{\sqrt{4\pi D(\Delta T)}} \exp \left( - \frac{((X_f+X_i)-\beta_x \Delta T +2nL_x)^2}{4D(\Delta T)} \right) 
\end{equation}

\begin{equation}
I_{2n} = \frac{1}{\sqrt{4\pi D(\delta T)}} \exp \left(  - \frac{(\Delta X -\beta_x \Delta T +2nL_x)^2}{4D(\Delta T) }
 \right)
\end{equation}
Analogous expressions, for Y - coordinates, $L_y$ dimension and $\beta_y$, correspond to $I'_{1n}, I'_{2n}$.

The model  provides now the  migration proportions $w_{if}$ defined by the fraction of paths which start in a given area ($A_i$) at time $T_0$ and are found in an area ($A_f$) after a time $T$. Let us denote $X_i^U$, $Y_i^U$, $X_i^L$, $Y_i^L$ and $X_f^U$, $Y_f^U$, $X_f^L$, $Y_f^L$ the coordinates of upper - right and lower - left corners of two rectangular areas $A_i$ and $A_f$ respectively.

The initial conditions on the initial large area are given by a  uniform distribution at time $T_0$. This is consistent with the long time limit of small scale solutions.

\vspace{0.3cm}
\noindent {\bf Proposition 2.1}

\noindent The proportions $w_{if}$ are given by:

\begin{equation} \label{proportions}
w_{if} = \frac{\int_{A_f} da_f  \int_{A_i} da_i G({\bf  R}_f - {\bf R}_i \mid \Delta T)}{\int_{\Omega} da_f  \int_{A_i} da_i G({\bf R}_f - {\bf R}_i \mid \Delta T)}
\end{equation}
which, due to (\ref{largeSol}) becomes: $w_{if} = w^x_{if} w^y_{if}$, with $w^x_{if} = n^x_{if}/n^x_{ii}$ and:

\begin{equation} \label{wx}
n^X_{if} = \int_{X_f^L}^{X_f^U} dX_f  \int_{X_i^L}^{X_i^U} dX_i \sum_n \left( I_{1n} + I_{2n} \right)
\end{equation}
Similar expressions can be written for $w^y_{if}$.

Each term in the sum (\ref{wx}) has a tractable form. We give as an example  $n^X_{if} = \sum_n I^n_{11} - I^n_{12} + I^n_{21} - I^n_{22}$, where:
\begin{eqnarray*} \label{largeIterms}
& I^n_{11} = \tilde F(X_f^U + X_i^U - \beta_x \Delta T + 2n L_x)- \tilde F(X_f^L + X_i^U - \beta_x \Delta T + 2n L_x) \\ \nonumber
& I^n_{22}= \tilde F(X_f^U + X_i^L - \beta_x \Delta T + 2n L_x)- \tilde F(X_f^L + X_i^L - \beta_x \Delta T + 2n L_x)  \\ \nonumber
& I^n_{12} = \tilde F(X_f^U - X_i^L - \beta_x \Delta T + 2n L_x)- \tilde F(X_f^L - X_i^L - \beta_x \Delta T + 2n L_x)  \\ \nonumber
& I^n_{21} = \tilde F(X_f^U - X_i^U - \beta_x \Delta T + 2n L_x)- \tilde F(X_f^L - X_i^U - \beta_x \Delta T + 2n L_x) 
\end{eqnarray*}
and: 
$
\tilde F (z)= F(\tilde z)=  \frac{1}{2}(\tilde z \cdot  erf(\tilde z) + \frac{1}{\sqrt{\pi}} \exp{(- \tilde z^2)})
$ 
for $\tilde z = \frac{1}{2 \sqrt{D \Delta T}} z$.

\noindent Similar formulae can be written for the $Oy$ terms, thus the proportions $w_{if}$ have closed expressions. Equation (\ref{proportions}) is just the definition of the probability of finding the final position in $A_f$, given that the initial position lies in $A_i$, normalized by $\int_{\Omega} da_f  \int_{A_i} da_i G({\bf R}_f - {\bf R}_i \mid \Delta T)$ since the solution of Focker - Planck equation does not necessarily integrate to 1 for arbitrary boundary conditions (\ref{BClarge}). The next steps above are just elementary calculations.

We will show in next section that there exist meaningful parameters for the drift and diffusion coefficient used in calculating the proportions $w_{if}$, and that they can be estimated consistently and without bias.

\section{Statistical model}

In this section, we will define the joint distributions of the {\it observed} positions. We then identify meaningful parameters of the resulted statistical model and find estimators which can be used for long time and large scale inference.

Due to the separability of the solutions in section 2.1, we can restrict the present calculations to one - dimensional case. One can easily check the validity of two - dimensional generalization of these results.

\subsection{The model parameters}

Under the stochastic model previously described (section 2.1), for any given $\delta t_i$, each $\delta x_i$ is normally distributed, with cumulants: $k_{1(i)} = \beta \delta t_i$, $k_{2(i)} = 2 D \delta t_i$ if $D$ is constant, or $k_{2(i)} = 2 \int_0^{\delta t_i}D(t) dt$ when $D$ depends on time.

The joint distribution of errors $ (\epsilon_1,  ...,  \epsilon_n)$ has a null vector as mean and diagonal higher order (tensor) cumulants, since the errors are i.i.d.

Each difference $\delta \epsilon_i$ has a distribution defined by the following cumulants: $k_{1(i)}^{\delta \epsilon} = 0$, $k_{2(i)}^{\delta \epsilon} = 2 k_2^{\epsilon}$, $k_{2p+1(i)}^{\delta \epsilon} =0$, $k_{2p(i)}^{\delta \epsilon} = 2 k_{2p}^{\epsilon}$, for $p = 1,2,...$. 

However, their joint distribution will  have non-diagonal higher cumulants. For example, $k_{2(i, i \pm 1)}^{\delta \epsilon} =  k_2^{\epsilon}$.

Although the joint multivariate distribution of $(\delta x_1, ... , \delta x_n)$ has a diagonal second order cumulant (the variance - covariance matrix) and all higher order (tensor-) cumulants are zero, this is not the case for the joint distribution of the observed values  $(\delta x_1^{obs}, ... , \delta x_n^{obs})$. The first cumulant of this distribution is the same vector which is the cumulant of the true (under the model) values, i.e. ${\bf k}_1 = (\beta \delta t_1, ..., \beta \delta t_n)$. The non-null elements of the variance - covariance matrix are given by $ k^{(obs)}_{2,(ii)} = k_{2(i)} + 2 k_2^{\epsilon} $, $ k^{(obs)}_{2,(i, i \pm 1)} = -  k_2^{\epsilon}+ (k_{2(i)} - k_{2(i \pm 1)})$. The elements of higher order cumulants depend only on the error distribution and can be straightforwardly calculated. An useful example is given in the following property.

\vspace{0.3cm}
\noindent {\bf Proposition 3.1}

\noindent The joint cumulants of the type $k_a(\delta x_i^{obs},...,\delta x_{j}^{obs}, ...)$, where $\delta x_i^{obs}$ appears $a_i$ - times and $\delta x_j^{obs}$ appears $(a-a_i)$ - times, when $a>2$ satisfy the following relations:

\begin{equation} \label{smallCumul}
k_a(\delta x_i^{obs},...,\delta x_{j}^{obs}, ...) = (1 + (-1)^a) \cdot  k_a^{\epsilon} 
\end{equation}
if $i=j$, $k_a(\delta x_i^{obs},...,\delta x_{j}^{obs}, ...) =(-)^{a_i} k_a^{\epsilon}$ for $j=i-1$, $k_a(\delta x_i^{obs},...,\delta x_{j}^{obs}, ...) =(-)^{a_{i+1}} k_a^{\epsilon}$ for $j=i+1$ and zero otherwise.
The proof is given in the Appendix.

\vspace{0.3cm}
\noindent {\bf Remark 3.1}

The available data consists of observations $ x_0^{obs}, ... ,x_n^{obs}$,  at $t_0, ..., t_n$, i.e. $\delta x_1^{obs}, ..., \delta x_n^{obs}$ corresponding to $\delta t_1, ...., \delta t_n$. The time intervals are thus distributed according to some (discrete) probability density (with weights $p(\delta t_i))$. This is however an advantage for our purposes, since it implies that, in a long time $\Delta T \rightarrow \infty$,  any given $\delta t_i$ is sampled $n_i$ - times, with $n_i = n p(\delta t_i)$, where $n$ is the total number of intervals $\sum_i \delta t_i = \Delta T$. Therefore, for each distinct $\delta t_i$, a number of $n_i$ values of $\delta x$ are sampled from a common distribution with first and second cumulant depending on $\delta t_i$.

As a consequence, for large enough values of $n_i$, one could estimate the individual mean and variance of each $\delta x_i^{obs}$  distribution. However, our goal being statistical inference over long time $\Delta T$, we need to estimate the parameters of the  $\Delta X = \sum_i \delta x_i^{obs}$ distribution. We give the theoretical expressions of these meaningful parameters in what follows.

\vspace{0.3cm}
\noindent {\bf Definition 3.1}:
For a given discretely observed (with noise) diffusion process over $\sum_i \delta t_i = \Delta T$, the {\it effective drift} parameter $\beta_{eff}$ is defined by:

\begin{equation}
\beta_{eff}  \sum_i \delta t_i = \sum_i k_{1(i)} = k_1(\Delta X)
\end{equation}

This implies:

\begin{equation} \label{betaEff}
\beta_{eff} \sum^*_i p(\delta t_i) \delta t_i = \sum^*_i p(\delta t_i) k_{1(i)}
\end{equation}
where the sums $\sum^*_i$ are over {\it distinct} values of $\delta t_i$.
In the continuum limit of the time interval distribution, the equation (\ref{betaEff}) becomes:

\begin{equation}
\beta_{eff} = \frac{\int  k_1[\delta t] p(\delta t) d(\delta t)}{\int  (\delta t) p(\delta t) d(\delta t)}
\end{equation}
We indicate the dependence on time interval of any cumulant by using the notation $k_a[\delta t]$.  

\vspace{0.3cm}
\noindent {\bf Definition 3.2}:
For a given discretely observed (with noise) diffusion process over $\sum_i \delta t_i = \Delta T$, the {\it effective diffusion} parameter $D_{eff}$ is defined by:

\begin{equation}
 2 D_{eff}  \sum_i \delta t_i =  k_2(\Delta X, \Delta X)
\end{equation}
where $\Delta X = \sum^n_i \delta x_i^{obs}$.
We can make explicit the second cumulant of $\Delta X$ distribution , in terms of $\delta x^{obs}_i$ - second cumulants:

\begin{equation} 
k_2(\Delta X, \Delta X) = \sum_i \sum_j k_2(\delta x_i^{obs}, \delta x_j^{obs})
\end{equation}
which after simple processing gives:

\begin{equation} 
k_2(\Delta X, \Delta X) = \sum_i  k_{2(i)} + 2k_2^{\epsilon} = \sum^*_i n_i k_{2(i)} + 2k_2^{\epsilon}
\end{equation}
where again, the sums $\sum^*_i$ are over {\it distinct} values of $\delta t_i$.
The first equality was obtained by using the properties of the joint distribution of $(\delta x_1^{obs}, ..., \delta x_n^{obs})$ mentioned at the beginning of this section:
$ \sum_i \sum_j k_2(\delta x_i^{obs}, \delta x_j^{obs}) =$ 

\noindent $  \sum^{n-1}_{i=2} \left( k_2(\delta x^{obs}_i, \delta x^{obs}_{i-1}) + k_2(\delta x^{obs}_i, \delta x^{obs}_i) +k_2(\delta x^{obs}_i, \delta x^{obs}_{i+1})    \right) + k_2(\delta x^{obs}_1, \delta x^{obs}_1) + k_2(\delta x^{obs}_1, \delta x^{obs}_2) + k_2(\delta x^{obs}_n, \delta x^{obs}_n) + k_2(\delta x^{obs}_n, \delta x^{obs}_{n-1}) =
   \sum^{n}_{i=1}  k_2(\delta x_i, \delta x_i) + 2k_2^{\epsilon}
$

\noindent The resulted equation for $D_{eff}$ is then:
\begin{equation} \label{DEff}
2 D_{eff} \sum^*_i p(\delta t_i) \delta t_i = \sum^*_i p(\delta t_i) k_{2(i)} + 2 n^{-1} k_2^{\epsilon}
\end{equation}
In the continuum limit of the time interval distribution (large $n_i$ and large $n$, small $\delta t$), the equation (\ref{DEff}) becomes:

\begin{equation}
D_{eff} = \frac{\int  k_2[\delta t] p(\delta t) d(\delta t)}{ 2 \int  (\delta t) p(\delta t) d(\delta t)}
\end{equation}
The error term in (\ref{DEff}) is $O(n^{-1})$, and, provided $k_2^{\epsilon}$ is finite, does not contribute to the continuous limit above.

\vspace{0.3cm}
\noindent {\bf Remark 3.2}

\noindent It is easy to check that if $D$ is constant or the distribution of the time intervals is uniform, the effective parameter coincides with the constant value or the integrated $\int D(t) dt$, respectively.

\vspace{0.3cm}
\noindent {\bf Remark 3.3}

\noindent The distribution of large scale values $\Delta X$ is not completely specified by the first two cumulants, and the corresponding Focker - Planck equation should contain higher order spatial derivatives. However, keeping only the first two terms is a reasonable approximation, since we can easily check that $k_3(\Delta X, \Delta X, \Delta X) = 0$, and only at the forth order we obtain $k_4(\Delta X, \Delta X, \Delta X, \Delta X) = 8 k_4^{\epsilon} (2n-1)$, thus a forth order derivative with a $8 k_4^{\epsilon}$ - coefficient in the generalized Focker - Planck equation.

In applications we can encounter the problem of observing many diffusion paths which are not necessarily generated by the same stochastic process, i.e. with the same drift and diffusion coefficients. For example, in ecological systems models, the paths correspond to different individuals which can have different behaviour. However, if the goal of statistical inference is long term and large space scale predictions for the {\it ensemble} of diffusion processes (the group of individuals), we can define new parameters which will describe this ensemble. We will call them {\it collective drift and diffusion coefficient}.

Let $\beta_{eff}^{\gamma}$, $D_{eff}^{\gamma}$, $\Delta T^{\gamma}$, $k_1(\Delta X)^{\gamma}$, $k_2(\Delta X, \Delta X)^{\gamma}$ be the parameters, characteristic long time and large distance and cumulants of any path $\gamma$ from an arbitrary set $\Gamma$.

\vspace{0.3cm}
\noindent {\bf Definition 3.3}

\noindent The {\it collective} drift and diffusion coefficient for the ensemble $\Gamma$ are defined by the equations:

\begin{equation}
\beta^{\gamma}_{collect} E_{\Gamma}(\Delta T^{\gamma}) = E_{\Gamma}(k_1(\Delta X)^{\gamma})
= E_{\Gamma}(\beta_{eff}^{\gamma} \Delta T^{\gamma})
\end{equation}

\begin{equation}
2 D^{\gamma}_{collect} E_{\Gamma}(\Delta T^{\gamma})= E_{\Gamma}(k_2(\Delta X, \Delta X)^{\gamma})
= 2 E_{\Gamma}(D_{eff}^{\gamma} \Delta T^{\gamma})
\end{equation}
where the expectations $E_{\Gamma}(f^{\gamma})$ are calculated over the ensemble of paths.

\subsection{The estimators}

We propose now consistent and unbiased estimators of the effective and collective parameters defined in the previous subsection. 

\vspace{0.3cm}
\noindent {\bf Proposition 3.2}

\noindent  For each distinct $\delta t_i$, let $ (\delta x_i^{obs}) ^{(\alpha)}$, $\alpha = 1,...,N$, be $N$  values sampled from a distribution described by its cumulants $k^{obs}_{a (i)}$, $a=1,2...$. Here, $N$ can be the actual number of available observations ($n_i$) or a number of values obtained by re-sampling with replacement from each such a group. Assume $k^{obs}_{1 (i)}=0$, to simplify notations. For non-null means, one needs to "center" the observations (subtracting the estimated means), but all properties remain {\it valid} in what follows if that is the case.

\noindent  Denote by $\overline {\delta x_i^{obs}}$ the sample average of $(\delta x_i^{obs}) ^{(\alpha)}$. The estimator of the effective drift parameter given by:

\begin{equation} \label{betaEffEst}
\hat \beta_{eff} = \frac{ \sum_i \overline {\delta x_i^{obs}} }{ \sum_i \delta t_i }
\end{equation}
is consistent and unbiased.

\vspace{0.3cm}
\noindent {\bf Proposition 3.3}

\noindent Under the same assumptions as in Proposition 2, a consistent and unbiased estimator of the effective diffusion parameter is given by:

\begin{equation}  \label{DEffEst}
\hat D_{eff} = \frac{ \sum_i  \sum_j \overline {\delta x_i^{obs} \delta x_j^{obs}} }{ 2 \sum_i \delta t_i }
\end{equation}

\noindent The prove of both propositions is given in the Appendix.

In a similar manner, averaging over paths will give estimators for the collective parameters, which in turn can be used in estimating the migration proportions as derived in section 2.2. Re-sampling (with replacement) methods can easily provide confidence intervals for either effective (when re-sampling observations of each path) parameters or collective ones (when re-sampling in two stages, at path and observation level).

\vspace{0.3cm}
\noindent {\bf Remark 3.4}

The relation between the observed large scale and true (under the model) cumulants is additive, so knowing the error variances $k_2^{\epsilon}$ allows us to determine the later from (\ref{DEff}) and (\ref{DEffEst}).

\vspace{0.3cm}
\noindent {\bf Remark 3.5}

\noindent 
The estimators we propose are fundamentally different from the ones derived in literature (see \cite{3} for a good review). Usually, the variance or the integrated variance, when time dependence is allowed) of the observed positions is usually obtained under the assumption of uniform time - interval distribution or, if not, by relying on Taylor expansions. In all these cases, the estimator (or its first approximation)  $\hat k_2(\Delta X, \Delta X)$ was of the type $ \left( \sum (\delta x_i^{obs})^2 \right/n$. We exploit the time interval distribution and the correlation structures in a different manner. In the particular case of uniform distribution one may easily check that we recover the known estimators.

\section{Case study}

In this section, we apply our proposed model to a real data set which consists of locations recorded at finite random time intervals by satellite tags attached to 19 hooded seals. The hooded seal ({\it Cystophora cristata}) is a key pinniped species in the Greenland and Norwegian Seas. 

The distribution and behaviour of these animals have been studied \cite{13},\cite{14}  by tagging a group of seals with satellite - linked platform terminal transmitters (PTT) on the sea ice near Jan Mayen. A total of 12,834 locations were determined during an overall tracking period of 3,787 seal days, and their range was very vast: from 54° N to 84° N, and from 41° W to 16° E. 

In figures 1, 2, 3, we give examples of the migration paths of 3 seals and the empirical distributions of: the distances along Ox (absolute value of longitude) and Oy (latitude) - axes, the lengths of the time intervals between measurements and the centered and scaled observations.

Two models were proposed for this data. An effective, individual model, which provided estimated diffusion parameters for each seal, and a collective one, which gives estimates of the parameters characterizing the whole group.

The models were also tested against each other, in order to decide which one is more appropriate for statistical inference. A test based on the asymptotic approximation for the distribution of these parameters gave non-significant differences between individual and collective parameters. The same conclusion is illustrated by the qq - plot in figure 5 which is  obtained from the empirical distributions under the two models generated by re-sampling with replacement. 

The main stochastic effect seems to be due to pure diffusion, the collective  drift is very weak ($(\beta \Delta T)^2 << 2 D \Delta T $, for $\Delta T$ of the order of 3, 4 or 6 months). This is in accordance to previous observations which indicate that hooded seals do not display any general seasonal migration pattern.

\section{Conclusions}

In this article, we have modeled two types of discrete data (observed at two space - time scales) by a unique diffusion process. This allowed us: (i) to derive consistent and unbiased statistical estimators for large scale model parameters as functions of small scale observations, but also (ii) to express large scale quantities of interest (like migration proportions) in terms of a minimal number of parameters.

We have applied this procedure to a migration data set, where only small scale discrete observations were available.

In addition, since the methods described here give a flexible
description of (although are not restricted to) migration processes and have closed-form solutions, they can be readily incorporated in complex models of population dynamics, such as models implemented in Gadget \cite{7} or similar modelling environments.

\section{Appendix}

\vspace{0.3cm}
{\bf Proof of Proposition 3.1}:
\vspace{0.3cm}

Let $\delta x_i^{obs} = x_{i}^{obs} - x_{i-1}^{obs}$, for $i=1,2,...n$, where $n$ is the number of samples.  Then $ \delta x_i^{obs} = x_{i} + \epsilon_i - x_{i-1} - \epsilon_{i-1}$. Let $x_i$ be distributed as (\ref{smallSol}) and let $\epsilon_i$ be i.i.d., i.e. $k_a(\epsilon_i, ..., \epsilon_i)= k_a^{\epsilon}$, where $k_a(\epsilon_i, ..., \epsilon_i)$ stands for the joint cumulant of $a$ terms $\epsilon_i$.

The joint cumulants $k_a( \delta x_i^{obs}, ....\delta x_j^{obs}, ...)$, where the total number of terms is $a$, can be calculated by using multi-linearity properties of cumulants. 
Any $x_i$ and $\epsilon_i$ are independently distributed, so any joint cumulants involving this type of terms cancel. The small scale distribution of the true positions is a Gaussian, so all higher cumulants ($a>2$) of the type $k_a( \delta x_i, ....\delta x_j)$ are zero.

Therefore, when $a>2$,we are left with terms of the type: $k_a(\epsilon_i - \epsilon_{i-1}, ..., \epsilon_j - \epsilon_{j-1}, ...)$. This can be written as:  $k_a(\epsilon_i, ..., \epsilon_j, ...)$   $+  (-)^{a_i} k_a( \epsilon_{i-1},  ..., \epsilon_j,  ...) +$   

\noindent $  (-)^{a_j}  k_a( \epsilon_i,  ..., \epsilon_{j-1},  ...)$  $ +  (-)^a  k_a( \epsilon_{i-1},  ...,  \epsilon_{j-1}, ...)$.

We denoted by $\epsilon_i, ...$ a number $a_i$ of $\epsilon_i$'s and similar for $j$, $i-1$ and $j-1$.

When $j=i$, only the first and last contributions are non-zero, so that:
$ = k_a^{\epsilon} + (-)^a k_a^{\epsilon}$. 
When $j=i-1$, only the second type of contributions do not cancel:
$ (-)^{a_i} k_a(\epsilon_{i-1}, ...,\epsilon_j, ...) = (-)^{a_i} k_a (\epsilon, ...\epsilon)= (-)^{a_i} k_a^{\epsilon}$ while for $j=i+1$ we obtain: $(-)^{a_{i+1}} k_a(\epsilon_{i}, ...,\epsilon_{i+1}, ...) = (-)^{a_{i+1}} k_a (\epsilon, ...\epsilon)= (-)^{a_{i+1}} k_a^{\epsilon}$.

\noindent Note:

\noindent For $a= 2$ we obtain $k_2(\delta x_i^{obs}, \delta x_i^{obs})= k_2(\delta x_i, \delta x_i) + 2 k_2^{\epsilon} = 2 \int_{t_{i-1}}^{t_i} D(t) dt + 2 k_2^{\epsilon}$ and $k_2(\delta x_i^{obs}, \delta x_j^{obs})= - k_2^{\epsilon} + 2 \int_{\delta t_{i \pm 1}}^{\delta t_i} D(t) dt $ for $j=i \pm 1$.

\vspace{0.3cm}
{\bf Proof of Proposition 3.2}:
\vspace{0.3cm}

(i) the estimator (\ref{betaEffEst}) is unbiased:

\noindent $ E(\hat \beta_{eff} \sum_i \delta t_i ) = E(  \sum_i \overline {\delta x_i^{obs}} ) =  \sum_i E(  \overline {\delta x_i^{obs}} ) =  \sum_i E (\frac{1}{N} \sum_{\alpha}  (\delta x_i^{obs})^{(\alpha)} ) = \sum_i k_{1(i)} = \beta_{eff} \sum_i \delta t_i $.

(ii) the estimator (\ref{betaEffEst}) is consistent:

\noindent $ \hat \beta_{eff} \sum_i \delta t_i = \sum \hat k_{1(i)}$ and (see \cite{15}) each $ \hat k_{1(i)}  \xrightarrow{p} k_{1(i)}$.

\vspace{0.3cm}
{\bf Proof of Proposition 3.3}:
\vspace{0.3cm}

(i) the estimator (\ref{DEffEst}) is unbiased.

\noindent  $ E(2 \hat D_{eff} \sum_i \delta t_i ) = E(  \sum_i  \sum_j \overline {\delta x_i^{obs} \delta x_j^{obs}}  ) = \sum_i  \sum_j E( \hat k_{2,(ij)}  ) =  \sum_i  \sum_j k_{2,(ij)} = 2 D_{eff} \sum_i \delta t_i $

(ii) the estimator (\ref{DEffEst}) is consistent:
  
\noindent $ 2 \hat D_{eff} \sum_i \delta t_i  =  \sum_i  \sum_j  \hat k_{2,(ij)} $ and (see  \cite{15} )  each $ \hat k_{2,(ij)} \xrightarrow{p} k_{2,(ij)} $.

\newpage

\begin{figure}[htbp]
\begin{center}
\resizebox{\columnwidth}{!}{\includegraphics[angle=270]{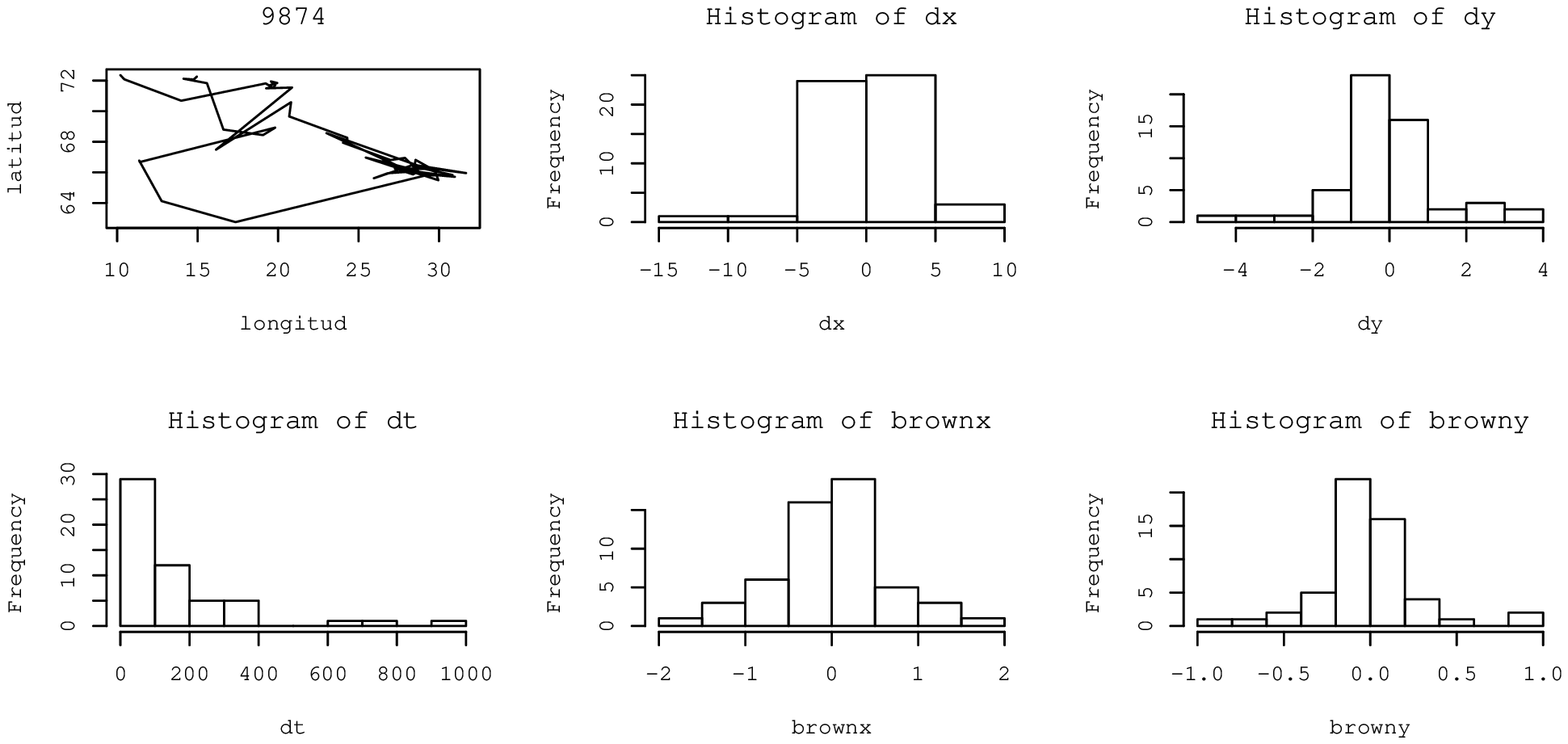}}
\caption{\small  Example of observed path (9874), empirical distributions of:  distances (dx and dy) along Ox and Oy respectively,  time  - intervals (dt) between observations, centered and scaled observations (brownx and browny respectively) along Ox and Oy (where Ox corresponds to absolute values of longitude and Oy - to latitude values).}
\end{center}
\end{figure}

\newpage

\begin{figure}[htbp]
\begin{center}
\resizebox{\columnwidth}{!}{\includegraphics[angle=270]{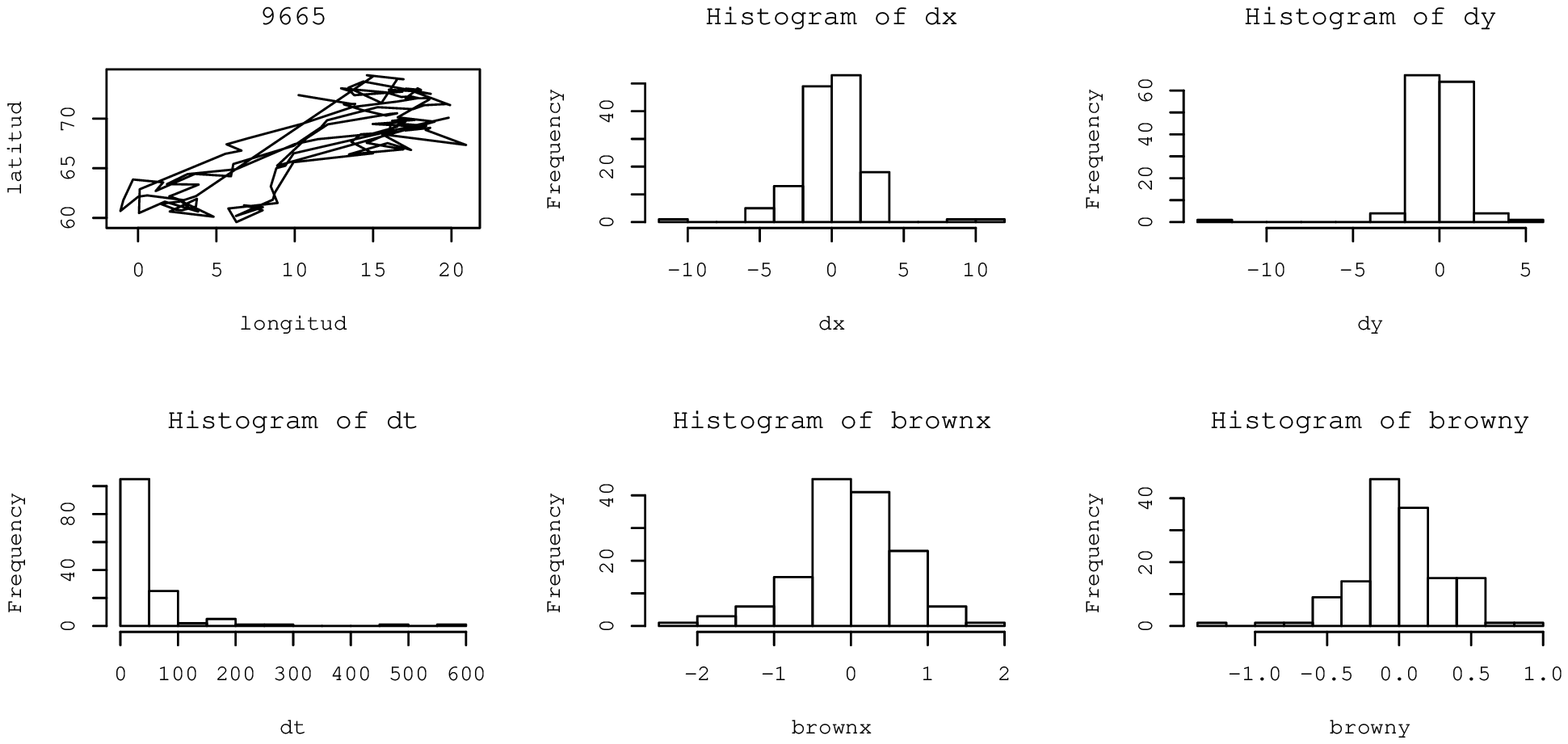}}
\caption{\small Example of observed path (9665), empirical distributions of:  distances (dx and dy) along Ox and Oy respectively,  time  - intervals (dt) between observations, centered and scaled observations (brownx and browny respectively) along Ox and Oy (where Ox corresponds to absolute values of longitude and Oy - to latitude values).}
\end{center}
\end{figure}

\newpage

\begin{figure}[htbp]
\begin{center}
\resizebox{\columnwidth}{!}{\includegraphics[angle=270]{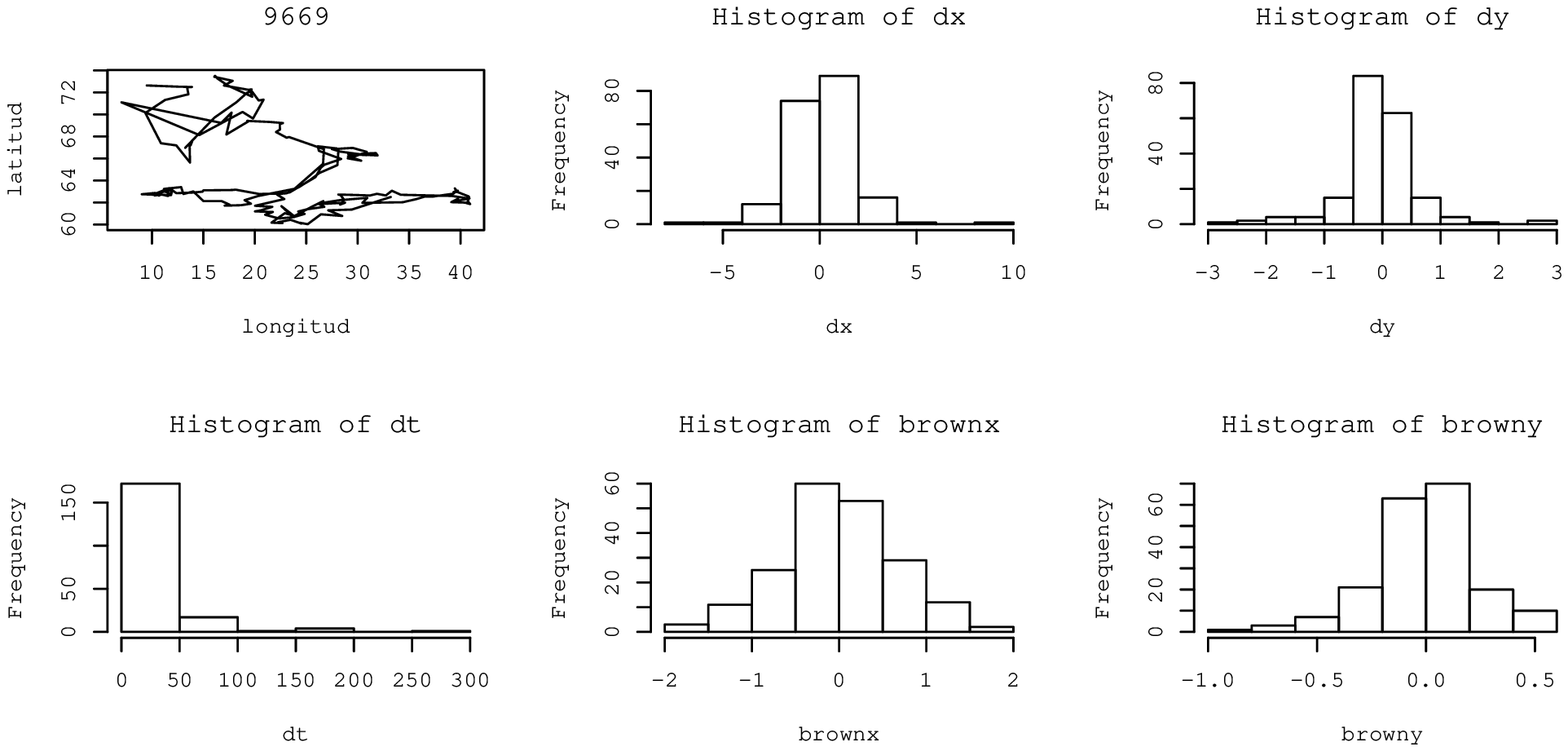}}
\caption{\small Example of observed path (9669), empirical distributions of:  distances (dx and dy) along Ox and Oy respectively,  time  - intervals (dt) between observations, centered and scaled observations (brownx and browny respectively) along Ox and Oy (where Ox corresponds to absolute values of longitude and Oy - to latitude values).}
\end{center}
\end{figure}

\newpage

\begin{figure}[htbp]
\begin{center}
\resizebox{\columnwidth}{!}{\includegraphics[angle=270]{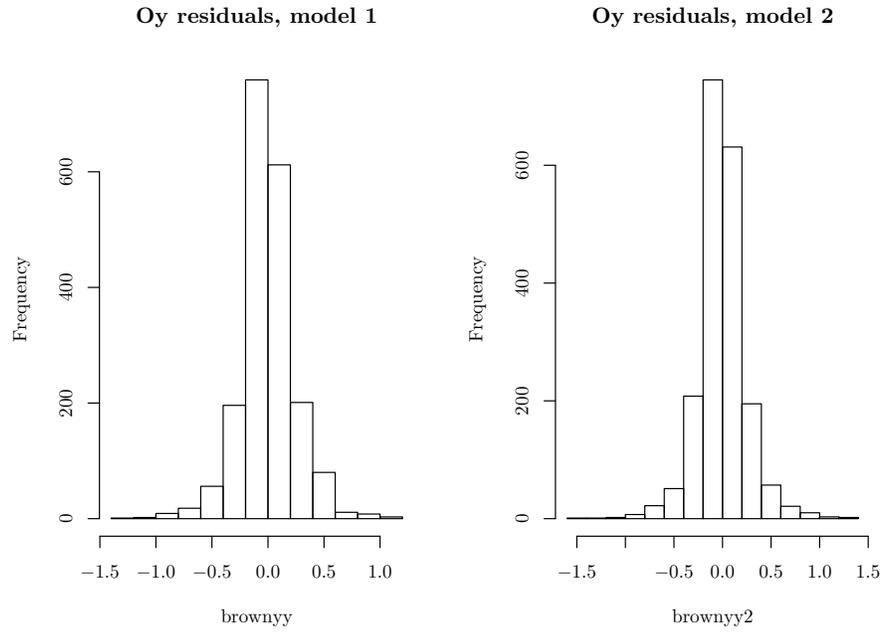}}
\caption{\small Empirical distributions of centered and scaled observations under effective model and collective model}
\end{center}
\end{figure}

\newpage

\begin{figure}[htbp]
\begin{center}
\resizebox{\columnwidth}{!}{\includegraphics[angle=270]{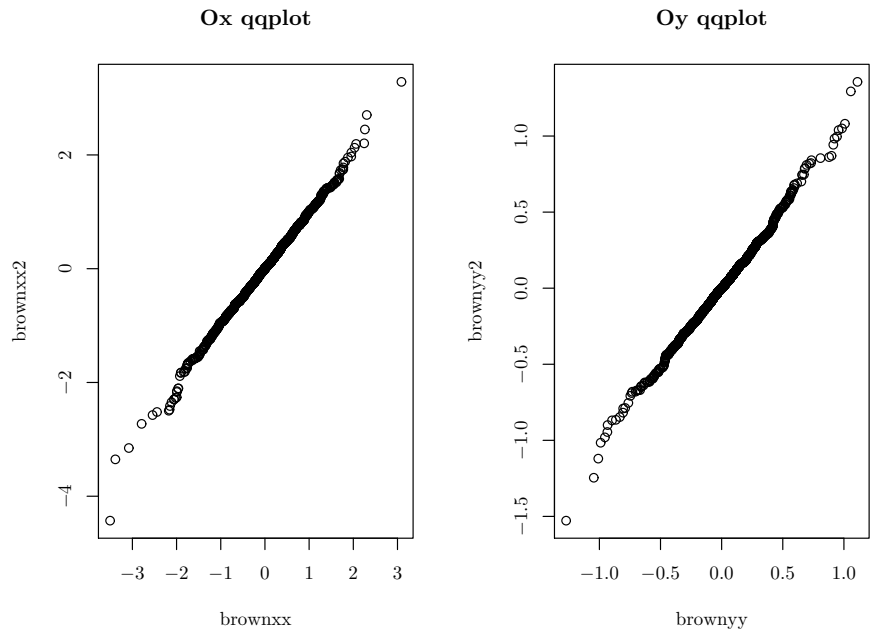}}
\caption{\small qq - plot testing collective versus effective models.}
\end{center}
\end{figure}

\end{document}